\def\kF{k_{\text{F}}}

\def\Ldw{L_{\text{dw}}}%
\def\Tc{T_{\text{c}}}

\def\gso{g_{\text{so}}}
\def\erf{{\text{erf}}}
\def\be{\begin{equation}}
\def\ee{\end{equation}}
\def\bea{\begin{eqnarray}}
\def\eea{\end{eqnarray}}
\def\bse{\begin{subequations}}
\def\ese{\end{subequations}}

\documentclass[prb,twocolumn,eqsecnum,showpacs,preprintnumbers,amsmath,amssymb]{revtex4}
\usepackage{graphicx}
\usepackage{dcolumn}
\usepackage{bm}
\usepackage{color}
\DeclareMathOperator\erfc{erfc}
\begin{document}
\preprint{}
\title{
        Stable phase separation and heterogeneity away from the coexistence curve}
\author{T.R. Kirkpatrick$^{1}$ and D. Belitz$^{2}$}
\affiliation{$^{1}$Institute for Physical Science and Technology, University of Maryland, College Park, MD 20742\\
                 $^{2}$Department of Physics, Institute of Theoretical Science, and Materials Science Institute, University
                           of Oregon, Eugene, OR 97403}
\date{\today}
\begin{abstract}
Phase separation, i.e., the coexistence of two different phases, is observed in many systems
away from the coexistence curve of a first-order transition, leading to a stable heterogeneous
phase or region. Examples include various quantum ferromagnets, heavy-fermion systems, 
rare-earth nickelates, and others. These observations seem 
to violate basic notions of equilibrium thermodynamics, which state that phase separation can
occur only on the coexistence curve. We show theoretically that quenched disorder allows for 
phase separation away from the coexistence curve even in equilibrium due to the existence
of stable minority-phase droplets within the majority phase. Our scenario also answers a related 
question: How can a first-order transition remain sharp in the presence of quenched disorder 
without violating the rigorous lower bound $\nu \geq 2/d$ for the correlation-length exponent?
We discuss this scenario in the context of experimental results for a large variety of systems.
\end{abstract}
\pacs{}
\maketitle

\section{Introduction}
\label{sec:I}

\begin{figure}[b]
\includegraphics[width=8.5cm]{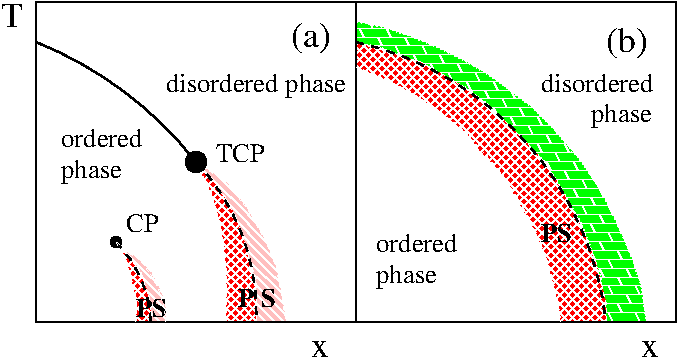}
\caption{Schematic phase diagrams in the temperature (T) - control parameter (x) plane. (a) A tricritical point (TCP)
              separates a line of second-order transitions (solid line) from a line of first-order transitions (dashed line).
              Phase separation (PS) is observed in both the cross-hatched (red) and hatched (pink) regions, although 
              some experiments show PS clearly only in the ordered phase. Examples are the helical magnet MnSi 
              \cite{Uemura_et_al_2007, Yu_et_al_2004} and the ferromagnet UGe$_2$ \cite{Harada_et_al_2005}, with 
              hydrostatic pressure as the control parameter. Inside the ordered phase another first-order transition and
              associated PS may be present, with the line of first-order transitions ending in a critical point (CP). An
              example is UGe$_2$ \cite{Kotegawa_et_al_2005}. (b) The transition is first order for all
              values of the control parameter. An example is the rare-earth nickelate Nd$_{1-x}$La$_x$NiO$_3$, with
              the dopant concentration as the control parameter \cite{Frandsen_et_al_2015, Uemura_2015}. PS in
              the tiled (green) region in the disordered phase is expected, but has so far not been observed.}
\label{fig:1}
\end{figure}

Phase separation, i.e., the coexistence of two different phases in thermodynamic equilibrium, is a
hallmark of first-order transitions. It follows from basic thermodynamics that this phenomenon can occur
only on the coexistence curve, where the two phases have the same free energy.\cite{Landau_Lifshitz_V_1980} 
However, in many solid-state systems phase separation is observed by a variety of techniques --~muon spin rotation 
($\mu$SR), nuclear magnetic resonance (NMR), nuclear quadrupole resonance (NQR), neutron depolarization
imaging, and neutron Larmor diffraction~-- even away from
a coexistence curve. Examples include quantum
ferromagnets and helimagnets such as MnSi,\cite{Uemura_et_al_2007, Pfleiderer_et_al_2010} 
Sr$_{1-x}$Ca$_x$RuO$_3$,\cite{Uemura_et_al_2007, Gat-Malureanu_et_al_2011} 
and UGe$_2$,\cite{Harada_et_al_2005, Kotegawa_et_al_2005}
heavy-fermion systems such as CeCu$_{2.2}$Si$_2$,\cite{Luke_et_al_1994} 
high-T$_{\text c}$ superconductors such as Europium-doped La$_{1.85}$Sr$_{0.15}$CuO$_4$,\cite{Kojima_et_al_2003} 
and systems displaying Mott transitions such as the rare-earth 
nickelates.\cite{Frandsen_et_al_2015, Frandsen_et_al_2016} In some of these systems the first-order transition
is from an ordered phase to a disordered phase (e.g., the ferromagnet-to-paramagnet transition in 
Sr$_{1-x}$Ca$_x$RuO$_3$,\cite{Gat-Malureanu_et_al_2011} 
or the transition from an antiferromagnetic insulator to a paramagnetic metal in the
nickelates \cite{Frandsen_et_al_2015}), in others, it is between two phases with the same order parameter (e.g., the FM1-FM2
transition between two ferromagnetic phases in UGe$_2$.\cite{Kotegawa_et_al_2005})
Some phase diagrams contain a tricritical point, in others the transition is first order for all accessible parameter
values, and in still others a line of first-order transitions ends in a critical point; schematic observed phase diagrams
are shown in Fig.~\ref{fig:1}. Phase separation is observed on either side of the transition, but the experimental
evidence is clearer in the ordered phase, and in some cases no phase separation has been observed so far in
the disordered phase. In some of these systems there is independent
evidence for the transition being first order, e.g., in the helical magnet MnSi,\cite{Pfleiderer_et_al_1997} in others
the observed phase separation is used as {\it prima facie} evidence for the first-order nature of a nearby phase
transition. 

These experiments and their interpretations raise a fundamental question: How can stable phase separation occur away from the
coexistence curve, where the two phases necessarily have different free energies? Surprisingly, this question does not
seem to have been addressed so far, although it is crucial for an understanding and interpretation of the experiments
mentioned above. We can think of only two possible
explanations: Either these systems are not in true thermodynamic equilibrium,\cite{Gunton_Miguel_Sahni_1983}
or they contain quenched disorder that couples to the order parameter and leads to the existence of static droplets 
of the minority phase within the majority phase. The first option
would require a non-equilibrium state with a very long relaxation time (at least years), since phase separation has
been observed in samples for which a first-order transition had been reported much earlier.\cite{Uemura_et_al_2007,
Pfleiderer_et_al_1997} While this is not inconceivable, it seems implausible
that such a state would not also lead to other observable consequences, e.g., glass-like features. We therefore focus on the
second possibility. 
We will show that a disorder-induced 
droplet scenario, namely, the existence of static minority-phase droplets within the majority phase, 
leads to a consistent theoretical picture that is in agreement with the experimental observations for a large class of 
systems.\cite{droplet_footnote} In particular, it can explain the deviation of the ordered-volume fraction from unity away from the coexistence
curve, and the observed asymmetry between the ordered and disordered phases.
It assumes the existence of quenched disorder that couples to the order parameter, but is not necessarily reflected
in transport experiments, as some of the systems in question are rather good metals. We will come back to this assumption
in the discussion.

The organization of this paper, and its main achievements, are as follows. In Sec.~\ref{sec:II} we use established 
results about first- and second-order transitions in the presence of quenched disorder to show that disorder-induced 
droplets can exist near first-order transitions, but not near second-order ones. In Sec.~\ref{sec:III} we establish criteria 
for the stability of minority-phase droplets. We show that they are stable in a sizable region of the phase diagram with
reasonable parameter values, and we estimate the ordered-volume fraction in the ordered phase as a function 
of the distance from the coexistence curve. In Sec.~\ref{sec:IV} we discuss our results and their relation to experiments
that have already observed the phenomena for which this paper provides a physical explanation.

\section{Phase Transitions in the Presence of Quenched Disorder}
\label{sec:II}

We start with some general considerations regarding second- and first-order transitions in the presence of
quenched disorder that couples to the order parameter. Let us first recall the Harris criterion for the critical 
behavior at a second-order transition to be unaffected by quenched disorder. Let $t$ be the dimensionless 
distance from the critical point, and $\nu$ the correlation-length exponent. Then the correlation length $\xi$ 
scales as $\xi \sim t^{-1/\nu}$, or $t \sim \xi^{-1/\nu}$.\cite{Landau_Lifshitz_V_1980, nu_footnote} Quenched disorder 
leads to an uncertainty in $t$. By the law of large numbers, this uncertainty will fall off as the inverse
square root of the system volume. Over a correlation volume, it thus obeys $(\Delta t)_{\text{dis}} \sim \xi^{-d/2}$,
where $d$ is the spatial dimensionality. In order for the transition to not be affected by the disorder, the
uncertainty $\Delta t$ must be smaller than $t$ itself. This leads to the condition \cite{Harris_1974}
\be
\nu \geq 2/d\ .
\label{eq:2.1}
\ee
In Harris's original argument \cite{Harris_1974} this condition referred to the renormalization-group fixed 
point that describes the critical point in a clean system. If it is violated, i.e., if $\nu < 2/d$ in the {\em clean}
system, then disorder either modifies the critical behavior or destroys the transition. No statement could
be made at the time about $\nu$ at the new fixed point, if any, that describes the transition in the presence of
disorder. Later, Chayes et al. \cite{Chayes_et_al_1986} proved rigorously that Eq.~(\ref{eq:2.1}) must hold at 
any fixed point that describes the physical critical behavior in the presence of quenched disorder.

Now consider a second-order transition in a finite system of linear size $L$ (which can be a subsystem of a
larger system) that is large compared to the microscopic length scale $a$, $L\gg a$. In the vicinity of the transition, the order-parameter susceptibility $\chi$ (among other observables)
obeys a homogeneity law \cite{Barber_1983}
\be
\chi(t,L) = b^{\gamma/\nu}\,F_{\chi}(t\,b^{1/\nu},L\,b^{-1}) = t^{-\gamma}\,F_{\chi}(1,L\,t^{\nu})\ .
\label{eq:2.2}
\ee
Here $b$ is an arbitrary scale factor, and $\gamma$ is the susceptibility exponent. The phase transition thus
gets rounded on a scale $(\Delta t)_{\text{rounding}} \sim L^{-1/\nu}$, while disorder fluctuations lead to an
uncertainty (see above) $(\Delta t)_{\text{dis}} \sim L^{-d/2}$. We now ask whether it is possible to form a
droplet of size $L$ that contains a distinguishable phase different from the majority phase. This 
requires $(\Delta t)_{\text{rounding}} < (\Delta t)_{\text{dis}}$, or 
\be
\nu < 2/d\ .
\label{eq:2.3}
\ee
This means that the existence of droplets is incompatible with the lower bound on $\nu$ given by Eq.~(\ref{eq:2.1}). We conclude
that disorder-induced droplets cannot exist in the vicinity of a second-order transition. 

Near a first-order transition, the situation is qualitatively different. As was shown by Fisher and Berker (Ref.~\onlinecite{Fisher_Berker_1982},
see also Ref.~\onlinecite{Privman_Fisher_1983}), finite-size scaling considerations yield 
\bse
\label{eqs:2.4}
\be
\nu = 1/d
\label{eq:2.4a}
\ee
for the correlation-length exponent at {\em any} thermal first-order transition. This has been generalized to
quantum phase transitions.\cite{Continentino_Ferreira_2004, Kirkpatrick_Belitz_2015b} It was shown that at $T=0$
Eq.~(\ref{eq:2.4a}) gets generalized to
\be
\nu = 1/(d + z_{\text{OP}})\ ,
\label{eq:2.4b}
\ee
\ese
where $z_{\text{OP}} > 0$ is the dynamical critical exponent that governs the temperature scaling of the order
parameter.\cite{Kirkpatrick_Belitz_2015b} The rounding scale thus crosses over from
$(\Delta t)_{\text{rounding}} \sim L^{-d} \ll L^{-d/2}$ in the classical case to
$(\Delta t)_{\text{rounding}} \sim L^{-(d+z_{\text{OP}})} \ll L^{-d/2}$ at $T=0$, and rounding does not preclude
the existence of droplets at any temperature. This striking difference between first- and second-order transitions gives a first indication 
of the physics behind the observed phase diagrams. 

To avoid misunderstandings we emphasize that the
droplets we consider are not rare regions. The length scale we consider is set by the correlation length,
so we are considering typical fluctuations rather then rare ones. We also stress that we are not 
discussing nucleation phenomena; the droplets we consider are static in nature. 

Let us now consider the energies relevant for the stability of droplets near a first-order phase transition. Consider a (sub)system
of linear size $L$ and volume $L^d$. Consider two states with free-energy densities $f_1$ and $f_2$, respectively. Suppose
a disorder fluctuation causes the lowest free-energy state in half of the system to be state 1, and in the other half, state 2.
The system can take advantage of this fluctuation by going into a phase-separated state, but this will incur an energy
cost in the form of a interface energy. For an order parameter with Ising symmetry, or for any $n$-vector model where
the coupling of the order parameter to the underlying lattice breaks the $O(n)$ symmetry, however weakly, the
interface energy is proportional to the surface area $L^{d-1}$. With $\sigma$ the surface 
tension, the free energy of the phase-separated state is
\bse
\label{eqs:2.5}
\be
F_{\text{ps}} = f_1\,L^d/2 + f_2\,L^d/2 + \sigma\,L^{d-1}\ ,
\label{eq:2.5a}
\ee
whereas the homogeneous state has a free energy
\be
F_{\text{hom}} = f_1\,L^d\ .
\label{eq:2.5b}
\ee
\ese
Now the free energy in a system with quenched disorder is a random variable, and hence its root-mean square
deviation is $\langle(\Delta F)^2\rangle^{1/2} \propto L^{d/2}$. If we take this to be representative of the 
free-energy difference $(f_1 - f_2)L^d$, then we have $(f_1 - f_2)L^d = 2\Delta\,L^{d/2}$, with $\Delta$ a measure
of the quenched disorder.\cite{Imry_footnote} The free-energy difference $\Delta F = F_{\text{ps}} - F_{\text{hom}}$
between the two phases then becomes
\be
\Delta F = -\Delta\,L^{d/2} + \sigma\,L^{d-1}\ .
\label{eq:2.6}
\ee
If $d<2$, we have $\Delta F < 0$ for any sufficiently large $L$. There thus is no energy barrier precluding the
existence of arbitrarily many droplets,
and the first-order transition will be smeared. However, if $d>2$, then the phase-separated state has a 
lower free energy than the homogeneous one only if $L$ is smaller than a critical value $L_c$ given by
\be
L_c = (\Delta/\sigma)^{2/(d-2)}
\label{eq:2.7}
\ee
For $d>2$, the largest possible linear droplet size is thus given by $L_c$.

The arguments related to Eqs.~(\ref{eqs:2.5}) - (\ref{eq:2.7}) are very similar to those given by Imry and 
Wortis,\cite{Imry_Wortis_1979} which in turn relied on a study of the random-field problem by Imry and
Ma.\cite{Imry_Ma_1975} These authors, and many others that used their ideas, focused on instabilities of 
phases with long-range order, and on the existence or otherwise of a sharp phase transition. While we
will comment on the latter aspect below, our main focus is
on a different aspect of the same arguments, namely, the idea that droplets of the ``wrong'' phase be
energetically stabilized inside the ``right'' phase by disorder fluctuations as long as their size does
not exceed a critical value. In this sense the two phases will coexist away from the coexistence
curve. The arguments say nothing about the region of the phase diagram where droplets
can be expected, and {\em a priori} it is not clear whether the allowable droplet size
makes them realizable. In what follows we will explore this scenario in more detail. Specifically,
we will explore how $L_c$ depends on the distance from the coexistence curve, what other
length scales are relevant for the problem, and what the resulting volume fraction of the majority
phase is expected to be. For definiteness, we will consider $d=3$, and we will consider droplets with no order
inside the ordered phase. We will discuss the issue of
ordered droplets within the disordered phase in Sec.~\ref{sec:IV}.

\section{Conditions for the Existence of Droplets}
\label{sec:III}

\subsection{Length scales, and probabilities}
\label{subsec:III.A}

There are various length scales in addition to the microscopic length $a$ that enter the problem, viz.: 
(1) The largest linear size a region or ``droplet'' favoring the minority phase can have and still occur 
with a probability that is not exponentially small. This we will denote by $L^*$; it is the largest size
of typical regions, as opposed to rare regions.
(2) The largest size a droplet can have and still be energetically favorable, taking into account both 
the energy gain due to the disorder fluctuation and the surface energy cost. This we will denote by 
$L_c$, as in Eq.~(\ref{eq:2.7}). 
(3) The minimum size a droplet must have in order to support an identifiable distinct phase. This we 
will denote by $L_0$. 
(4) The thickness of the droplet wall, which is important for determining the surface tension $\sigma$. 
This we will denote by $L_{\text{dw}}$. 
(5) In systems with a weakly broken continuous symmetry there are Goldstone modes. Let their
frequency-momentum relation in the long-wavelength limit be $\Omega(k) = s\,k^{\omega}$, with
$s$ a stiffness parameter. The breaking of the symmetry gives the Goldstone modes a gap
$\Omega_g$, which corresponds to a length scale $L_g = (s/\Omega_g)^{1/\omega}$.\cite{dynamics_footnote}
In an Ising-like system, $L_g \approx a$.

In order to discuss and estimate these characteristic length scales, consider a field theory with a random-mass
term, i.e., an action whose Gaussian part reads \cite{action_footnote}
\be
{\cal A}^{(2)} = \int_V d{\bm x}\,\phi({\bm x})\left[r + \delta r({\bm x}) - c{\bm\nabla}^2\right]\phi({\bm x})\ .
\label{eq:3.1}
\ee
Here $\phi({\bm x})$ is the order-parameter field. For simplicity, we consider a scalar order parameter,
we will consider the $n$-vector case below.
$V$ is the system volume, and $r$ and $c$ are parameters of the Landau-Ginzburg-Wilson functional ${\cal A}^{(2)}$.
$\delta r({\bm x})$ is a random variable governed by a distribution with zero mean and second moment
\be
\langle\delta r({\bm x})\,\delta r({\bm y})\rangle = \rho\,\delta({\bm x}-{\bm y}) \ ,
\label{eq:3.2}
\ee
which defines $\rho$. For simplicity, we assume $\delta r$ to be delta-correlated; we will come back to this assumption below.
If we take the order parameter to be dimensionless, then $r$ will be an energy density. Let $J$ be the
energy scale relevant for the order described (at a thermal phase transition, $J$ will be on the order of
the transition temperature $T_{\text{c}}$). Then we expect, up to
dimensionless factors, $c \approx J/a$, $r \propto J/a^3$, and
\be
\rho \approx \delta\,J^2/a^3\ ,
\label{eq:3.3}
\ee
where $\delta$ is a dimensionless measure of the disorder. Weak disorder corresponds to $\delta \ll 1$, and very strong
disorder corresponds to $\delta \approx 1$. Here, and throughout our discussion, we ignore dimensionless factors that
qualitative arguments give no control over.\cite{disorder_strength_footnote}

\subsubsection{The length scale $L^*$}
\label{subsubsec:III.A.1}

In order to estimate the length scale $L^*$, we consider $\delta r$ coarse-grained over a volume $L^3$ by defining
\be
\delta r_L := \frac{1}{L^3} \int_{L^3} \delta r({\bm x})\ .
\label{eq:3.4}
\ee
$\delta r_L$ is an average of independent random variables, so by the central limit theorem it is Gaussian
distributed,
\bse
\label{eqs:3.5}
\be
{\cal P}(\delta r_L) = \frac{1}{\sqrt{2\pi} s}\,e^{-(\delta r_L)^2/2s^2}
\label{eq:3.5a}
\ee
with second moment
\be
s^2 = \frac{\rho}{L^{6}} \int_{L^3} d{\bm x}\,d{\bm y}\ \delta({\bm x}-{\bm y}) = \rho/L^3\ .
\label{eq:3.5b}
\ee
\ese
Suppose the system undergoes a first-order transition at $r = r_1$. Let the system be in the 
ordered phase at $r < r_1$, and let $t = (r_1 - r) a^3/J$ be the dimensionless distance from
the coexistence curve. Then the probability of finding a region of size $L$ around any given point
that favors the disordered phase, i.e., where locally $\delta r_L> t$, is
\bse
\label{eqs:3.6}
\be
P_{\delta r_L>t} = \int_t^{\infty} dr_L\,{\cal P}(\delta r_L) = \frac{1}{2}\,\erfc\left((L/L^*)^{3/2}\right)\ .
\label{eq:3.6a}
\ee
Here $\erfc(x) = 1 - \erf(x)$ is the complementary error function, and
\be
L^* \approx a\,\delta^{1/3}/t^{2/3}\ .
\label{eq:3.6b}
\ee
\ese
where we have omitted a factor of $O(1)$. $L^*$ is the largest linear size a region or ``droplet'' favoring the
disordered phase can have and still be found with a probability that is not exponentially small. 

\subsubsection{The length scale $L_c$}
\label{subsubsec:III.A.2}

Now consider the disorder-induced contribution to the free energy. With $\varphi$ a characteristic value of the dimensionless order
parameter it is expected to be $F_{\text{dis}} \approx \overline{\delta r_L}\,\varphi^2 L^3$, with $\overline{\delta r_L}$
a characteristic value of $\delta r_L$. For the latter we take
\be
\overline{\delta r_L} = \int_t^{\infty} d(\delta r_L)\,\delta r_L\,{\cal P}(\delta r_L) \approx \rho\,L^{-3/2}\,e^{-\frac{1}{2}(L/L^*)^3}\ ,
\label{eq:3.7}
\ee
which is the average of $\delta r_L$ under the probability distribution ${\cal P}$ times the probability of finding a
droplet of this size. We thus have
\bse
\label{eqs:3.8}
\be
F_{\text{dis}} \approx \Delta(L)\,L^{3/2}
\label{eq:3.8a}
\ee
where
\be
\Delta(L) = \Delta_0\,e^{-\frac{1}{2}(L/L^*)^3}
\label{eq:3.8b}
\ee
with
\be
\Delta_0 = \delta^{1/2}\,J\,\varphi^2/a^{3/2}\ .
\label{eq:3.8c}
\ee
\ese
This is a generalization of the first term in Eq.~(\ref{eq:2.6}).

In addition to $F_{\text{dis}}$, we need to consider the surface energy cost. Suppose the order, upon approaching
a droplet, is gradually destroyed over a length $L_{\text{dw}}$, and the droplet wall is locally in the $y$-$z$
plane. Then the surface energy from the gradient-squared term in Eq.~(\ref{eq:3.1}) is 
$F_{\text{dw}} \approx L_y L_z c\varphi^2/L_{\text{dw}}$. The LGW coefficient $c$ is roughly $c \approx J/a$, and
hence the surface tension $\sigma = F_{\text{dw}}/L_y L_z$ is
\be
\sigma \approx J\varphi^2/a L_{\text{dw}}\ .
\label{eq:3.9}
\ee
The free-energy difference from Eq.~(\ref{eq:2.6}) now becomes
\be
\Delta F = -F_{\text{dis}} + F_{\text{dw}} = -\Delta(L) L^{3/2} + \sigma\,L^2\ ,
\label{eq:3.10}
\ee
and the largest possible droplet size $L_c$ is the solution of the transcendental equation
\bse
\label{eqs:3.11}
\be
L_c = L_c^0\,e^{-(L_c/L^*)^3}\ ,
\label{eq:3.11a}
\ee
which generalizes Eq.~(\ref{eq:2.7}). Here
\be
L_c^0 = (\Delta_0/\sigma)^2 \approx \delta L_{\text{dw}}^2/a\ .
\label{eq:3.11b}
\ee
Apart from logarithmic corrections in the case $L_c^0 \gg L^*$, the solution of Eq.~(\ref{eq:3.11a}) is
\be
L_c \approx \text{Min}(L_c^0,L^*)\ .
\label{eq:3.11c}
\ee
\ese

\subsubsection{The length scale $L_0$}
\label{subsubsec:III.A.3}

The length scale $L_0$ is determined by the condition that the energy gain $F_{\text{dis}}$ from 
the disorder fluctuation be larger than the characteristic energy $J$:\cite{Privman_Fisher_footnote}
\bse
\label{eqs:3.12}
\be
\Delta(L_0)\,L_0^{3/2} = J\ .
\label{eq:3.12a}
\ee
Using Eqs.~(\ref{eqs:3.8}) we have a transcendental equation for $L_0$,
\be
L_0 = L_0^0\,e^{(L_0/L^*)^3/3}\ ,
\label{eq:3.12b}
\ee
where
\be
L_0^0 = a/\delta^{1/3}\,\varphi^{4/3}\ .
\label{eq:3.12c}
\ee
This has a solution only if $L_0^0 \ll L^*$, in which case
\be
L_0 \approx L_0^0\ .
\label{eq:3.12d}
\ee
\ese

\subsubsection{The length scale $\Ldw$}
\label{subsubsec:III.A.4}

Of the length scales discussed so far, $L^*$ depends explicitly on the distance $t$ from the coexistence curve,
$L_0$ depends on $t$ only via $\varphi$, and $L_c^0$ depends on $t$ via $\Ldw$. In order to
estimate the latter, we must distinguish between scalar and vector order
parameters. Let us first discuss the latter case. 

A detailed determination of the droplet-wall thickness $\Ldw$ is a very hard problem, but we can gain sufficient 
insight from the case of a domain wall between ordered domains, say, ferromagnetic domains for the sake of
definiteness.\cite{Kittel_1996, Landau_Lifshitz_VIII_1984} In the case of a truly isotropic order parameter the
domain-wall width is equal to the system size. The reason is that an order-parameter modulation with an
infinite wavelength does not cost any energy; i.e., it is a consequence of the existence of gapless Goldstone
modes. In any real solid the underlying lattice couples to the order parameter and produces a small gap
$\Omega_g \ll J$, which corresponds to a length scale $L_g \gg a$, see Sec.~\ref{subsec:III.A}. The domain-wall
width in an infinite system is then given by 
\be
\Ldw \approx L_g/\varphi\ .
\label{eq:3.13}
\ee
In the case of a ferromagnet, $L_g \approx a/\gso$,\cite{Kittel_1996, Landau_Lifshitz_VIII_1984} with $\gso \ll 1$
the dimensionless spin-orbit coupling.

In our case we are interested in a wall between the ordered bulk and a disordered droplet, rather than one between
two ordered domains. This makes the problem more complicated, since the suppression of the long-ranged order
can occur via a loss of angular correlation, or via a modulation of the modulus of the order parameter, or both, but
we expect Eq.~(\ref{eq:3.13}) to still give the correct order of magnitude.\cite{Ldw_footnote}

We stress that the above considerations apply to an infinite system, or one whose linear
size $L$ is large compared to $L_g$. For $L<L_g$ one has $\Ldw \approx L/\varphi$, the
surface tension $\sigma$, Eq.~(\ref{eq:3.9}), is proportional to $1/L$, and the corresponding
contribution to the free energy scales as $F_{\text{dw}} \sim L^{d-2}$ rather than $L^{d-1}$.\cite{Imry_Wortis_1979} 
For our phase-separation problem this implies an intrinsic and
profound asymmetry between the ordered and disordered phases, as in the latter the size
of the ordered ``system'' is the droplet size. This implies that it is harder to form droplets in
the disordered phase. We will come back to this point in Sec.~\ref{sec:IV}.

Now consider making the anisotropy stronger and stronger. $\Omega_g$ will increase, and
$L_g$ will decrease, with the limiting case being $\Omega_g \approx J$, $L_g \approx a$,
for a strongly broken rotational symmetry. This limiting case describes an Ising system, as
can be checked by solving the saddle-point equation for an Ising-type action, see Appendix \ref{app:A}.
We conclude that Eq.~(\ref{eq:3.13}) is generally valid.
Deep inside the ordered phase, where $\varphi \approx 1$, we have $\Ldw \approx a$ for
Ising systems and $\Ldw \approx a/g \gg 1$, with $g$ the symmetry-breaking parameter,
for systems with a weakly broken rotational symmetry. (In the case of a ferromagnet,
$g = \gso$.) Near the coexistence curve, $\Ldw$ gets enhanced by a factor of $1/\varphi_1$,
with $\varphi_1$ the discontinuity of the order parameter at the first-order transition.

\subsubsection{Conditions for the existence of droplets}
\label{subsubsec:III.A.5}

We now are in a position to discuss the region in the phase diagram where droplets can exist.
The basic requirement is that the largest droplets that are energetically allowed are
large enough to host a distinguishable minority phase, i.e., we must have $L_c > L_0$. Since
$L_c$ is the smaller of $L_c^0$ and $L^*$, see Eq.~(\ref{eq:3.11c}), we need to distinguish
two cases: 

(i) If $L^* > L_c^0$, which is equivalent to $t < t^*$, with
\be
t^* = (a/L_g)^3 \varphi^3/\delta\ ,
\label{eq:3.14}
\ee
we have $L_c = L_c^0$, which leads to 
\bse
\label{eqs:3.15}
\be
\delta > \varphi_1^{1/2} (a/L_g)^{3/2} \equiv \delta_c\ .
\label{eq:3.15a}
\ee
Here we have replaced $\varphi$ by $\varphi_1$, the value of the order parameter on the
coexistence curve, since our considerations are valid close to a first-order transition.

(ii) If $L_c^0 > L^*$, which is equivalent to $t > t^*$, we have $L_c = L^*$,
which leads to 
\be
t < \delta\,\varphi_1^2 \equiv t_c
\label{eq:3.15b}
\ee
\ese
and we have again replaced $\varphi$ by $\varphi_1$. 
In this second case droplet thus can exist in a window of $t$-values, $t^* < t < t_c$,
and in order for this window to exists the inequality (\ref{eq:3.15a}) must again hold.

The Eqs.~(\ref{eqs:3.15}) are the two conditions
for the existence of droplets containing the disordered phase within the ordered phase.\cite{droplet_asymmetry_footnote}
Equation (\ref{eq:3.15a}) sets a threshold
for the disorder strength required for droplets to exist. It is smaller for more isotropic systems
(larger $L_g$), and for more weakly first-order transitions (smaller $\varphi_1$). For a given disorder strength, 
Eq.~(\ref{eq:3.15b}) determines the region
in the phase diagram where droplets can exist. The size of this region goes to zero for $\varphi_1 \to 0$,
consistent with our conclusion in Sec.~\ref{sec:II} that no droplets can exist in the vicinity of a
continuous transition. Now consider a phase diagram that includes a tricritical point,
such as in Fig.~\ref{fig:1}(a), and consider a point on the coexistence curve. Let us assume that
Eq.~(\ref{eq:3.15a}) is satisfied. This will require a very small amount of disorder for systems where
$L_g \gg a$, but is more difficult to fulfill in Ising-type systems; for an estimate within a very simple
model, see Appendix \ref{app:B}. Close to the tricritical point 
$\varphi_1$ is small, and droplets can exist only in a very small $t$-region according to
Eq.~(\ref{eq:3.15b}). As we move to lower temperatures, $\varphi_1$ increases, and reaches its
maximum at $T=0$. Equation (\ref{eq:3.15b}) thus predicts that droplets can exist in
a wedge-shaped region that emanates from the tricritical point. Analogous considerations hold
for phase diagrams where the tricritical point is not accessible (Fig.~\ref{fig:1}(b)), or where the
coexistence line ends in a critical point, Fig.~\ref{fig:1}(a). All of this is qualitatively consistent
with the observations, see the schematic phase diagrams shown in Fig.~\ref{fig:1}. 

\subsection{The ordered-volume fraction}
\label{subsec:III.B}

We now return to the probability considerations of Sec.~\ref{subsubsec:III.A.1} in order to obtain
an expression for the ordered-volume fraction. We start by considering the following conditional
probability: Given a disorder fluctuation that favors the paramagnetic phase, the probability density
of that region having a volume $V$ is, according to Eq.~(\ref{eq:3.6a}), 
\be
{\cal P}_{\delta r_L>t} = \erfc(\sqrt{V/V^*})/\int_0^{\infty} dV\, \erfc(\sqrt{V/V^*})\ ,
\label{eq:3.16}
\ee
with $V^* = (L^*)^3$. Of these regions, only those with volumes between $V_0 = (L_0)^3$
and $V_c = (L_c)^3$ have a lower free energy than the ordered phase once the surface 
energy is taken into account. The probability that a given disorder fluctuation forms a
minority-phase droplet is thus
\bse
\label{eqs:3.17}
\bea
P_d &=& \int_{V_0}^{V_c} dV\,{\cal P}_{\delta r_L>t}(V) 
\nonumber\\
       &=& p_d({\sqrt{V_c/V^*}}) - p_d({\sqrt{V_0/V^*}})\ ,       
\label{eq:3.17a}
\eea
where we have defined
\be
p_d(x) := \erf(x) + 2 x^2 \erfc(x) - 2x e^{-x^2}/\sqrt{\pi}\ .
\label{eq:3.17b}
\ee
\ese

\begin{figure}[t]
\includegraphics[width=8.5cm]{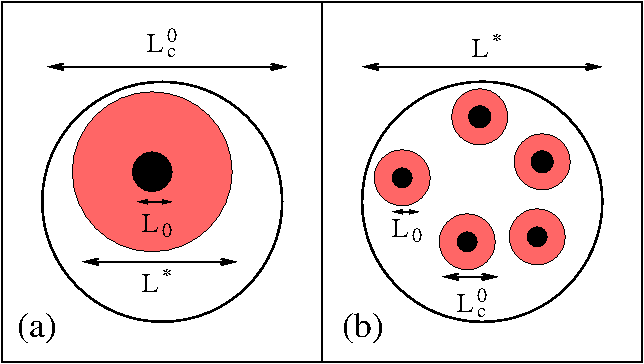}
\caption{(a) For $L_c = L^* < L_c^0$ each disorder fluctuation of size $L^*$ can support one droplet
               (orange (grey) circle) of size $L_c = L^*$. The minimum droplet size is $L_0$ (black
               circle), and droplets are expected to take up a volume fraction $1 - V_0/V^*$.
               (b) For $L^* \gg L_c^0 = L_c$ a disorder fluctuation can support on the order of
               $N_d \approx C_d V^*/V_c$ droplets, with $C_d$ a number that does not exceed the
               close-packing fraction for droplets. Droplets are expected to take up a volume
               fraction $C_d[1 - V_c^0/V_0]$.}
\label{fig:2}
\end{figure}

In the region $L^* < L_c^0$, where each disorder fluctuation can support only one droplet, this
leads to an ordered-volume fraction $F_{\text{OV}} \approx 1 - \frac{1}{2}\,P_d$. The factor of 
$1/2$ accounts for the conditional-probability nature of our starting point,
Eq.~(\ref{eq:3.16}): On the coexistence curve half of all disorder fluctuations will favor the 
disordered phase, and the other half will favor the ordered phase. In the region $L^* < L_c^0$,
which corresponds to $t > t^*$, we have, from Eq.~(\ref{eq:3.11c}), $L_c \approx L^*$, and hence
\bse
\label{eqs:3.18}
\be
F_{\text{OV}} \approx 1 - \frac{1}{2}\,\left[p_d(1) - p_d({\sqrt{V_0/V^*}})\right]\ , \quad (L^* < L_c^0)\ .
\label{eq:3.18a}
\ee
For $L^* > L_c^0$ the relation between $F_{\text{OV}}$ and $P_d$ changes. This is obvious for 
$L^* \gg L_c^0$, when a single disorder fluctuation
can contain many droplets. We roughly expect $F_{\text{OV}} \approx 1 - \frac{1}{2}\,P_d\,N_d$,
where $N_d$ is the number of droplets per volume $V^*$. Ignoring droplet-droplet interactions and a
factor of $O(1)$, we have $N_d \approx C_d\,V^*/V_c$, where $C_d$ is a number that is bounded above
by the close-packing volume fraction of the droplets within the volume $V^*$. The precise value of $C_d$
depends on many details; we expect it to be on the order of $1/2$. Since $L_c = L_c^0$ in this region,
this leads to an ordered-volume fraction
\bea
F_{\text{OV}} &\approx& 1 - \frac{C_d}{2}\,\left[p_d(\sqrt{V_c^0/V^*}) - p_d(\sqrt{V_0/V^*})\right]\,\frac{V^*}{V_c^0}\ ,
\nonumber\\
&&\hskip 100pt (L_c^0 < L^*)\ .
\label{eq:3.18b}
\eea
\ese
We note again that these simple arguments ignore any
interaction between the droplets as well as correlations between the disorder fluctuations and are
expected to give only a very rough estimate of the ordered-volume fraction. 
With this in mind, let us consider the behavior of $F_{\text{OV}}$ as a function of $t$ as the
coexistence curve is approached:
\smallskip\par\noindent
(1) $t > t_c$. In this region $L_c = L^* < L_0$. Equation~(\ref{eq:3.15b}) is 
violated, droplets cannot exist, and $F_{\text{OV}} = 1$.
\smallskip\par\noindent
(2) $t_c > t > t^*$. In this region $L_0 < L_c = L^*$. 
Droplets can exist, and $P_d$ increases from $0$ at $t = t_c$, where $V^* = V_0$, to 
$p_d(1) - p_d((\delta_c/\delta)^2)$ for $t = t^*$, with $\delta_c$ from Eq.~(\ref{eq:3.15a}). 
$p_d(1) \approx 0.74$, and $p_d((\delta_c/\delta)^2) \ll 1$ unless the disorder strength is close to the 
threshold value $\delta_c$. More specifically, $p_d(x\to 0) = 2x^2$, so for $V_0 \ll V^*$ we have
$P_d = p_d(1) - 2V_0/V^* = p_d(1) - 2(\delta_c/\delta)^2$, consistent with what one would expect 
from a simple geometric argument, viz., $P_d \approx 1 - V_0/V^*$, see Fig.~\ref{fig:2}(a). 
For $t = t^*$ we thus expect $F_{\text{OV}}$ to be on the order of $2/3$. 
\smallskip\par\noindent
(3) $t ^* > t$. In this region $L^*$ continues to increase and eventually multiple droplets will form within
each disorder fluctuation. In the limit $L^* \gg L_c^0$,  asymptotically close to the coexistence curve,
Eq.~(\ref{eq:3.18b}) yields $F_{\text{OV}} \approx 1 - C_d [1 - V_0/V_c^0] = 1 - C_d [1 - (\delta_c/\delta)^4]$. This is again consistent with
a the expectation from a simple geometric consideration, see Fig.~\ref{fig:2}(b).

\bigskip\par
In summary, for $t \agt t^*$ our mechanism results in behavior that is qualitatively consistent with the
experimental observations: Droplets can exist within certain distance $t_c$ from the
coexistence curve, and the ordered-volume fraction decreases from unity as the first-order transition
is approached from within the ordered phase. We will discuss the corresponding behavior in the disordered
phase in Sec.~\ref{sec:IV}.

\section{Discussion}
\label{sec:IV}
	
We conclude by discussing our results and their underlying assumptions in more detail, and
also add some remarks about aspects of the problem that we have not covered so far.

Before going into details, let us first reiterate how wide-spread the observations are. Phase separation is observed
near phase transitions that are known or suspected to be first order in a wide variety of
materials with an equally wide variety of types of order; examples were given in the
introduction. In addition to transitions from an ordered phase to a disordered one it also
is observed near first-order transitions from one ordered phase to another; an example
is UGe$_2$, where it is observed both near the FM1-PM transition \cite{Harada_et_al_2005} and
near the FM1-FM2 transition within the FM phase.\cite{Kotegawa_et_al_2005} Furthermore,
the observations are the same for systems with a strongly uniaxial or Ising-like order parameter,
such as UGe$_2$, and systems with an order parameter that is approximately rotationally
invariant or Heisenberg-like, such as MnSi.

As we pointed out in Sec.~\ref{sec:I}, these observations of the coexistence of two phases {\em away from
the coexistence curve} are very surprising and require an explanation. As we have shown,
disorder fluctuations provide a plausible scenario. We reiterate that the conditions for this scenario
to work require less disorder for systems with an approximate rotational symmetry than for
Ising-like ones, see the discussion after Eqs.~(\ref{eqs:3.15}), and Appendix \ref{app:B}. At the same
time, there is some experimental evidence for inhomogeneities that do not necessarily show in transport
experiments and therefore can be present even in nominally rather clean systems.\cite{Yu_et_al_2004}

\subsection{The role of disorder}
\label{subsec:IV.A}

We emphasize that our scenario requires a substantial amount of disorder that couples to
the order parameter. This is not to say that the disorder is necessarily visible in the
transport properties of metallic systems. If it were, then a rough estimate for our disorder parameter $\delta$
would be $1/\kF\ell$, with $\kF$ the Fermi wave number and $\ell$ the elastic mean-free path.
In clean samples of, e.g., MnSi, $\kF\ell$ can exceed $1,000$.\cite{Brando_et_al_2016} However,
Yu et al. have reported evidence for substantial pressure inhomogeneities
in MnSi.\cite{Yu_et_al_2004} The latter do indeed couple to the order parameter, as is evidenced by the strong
dependence of the transition temperature on applied hydrostatic pressure. Pfleiderer et al. 
have emphasized the sensitivity of systems near quantum phase transitions in general to 
disorder.\cite{Pfleiderer_et_al_2010} This can be illustrated as follows. 

Consider a thermal transition with a transition temperature
$\Tc$ that depends on a defect concentration $n({\bm x}) = n + \delta n({\bm x})$: 
\be
\Tc({\bm x}) \approx \Tc(n) + \delta n\, d\Tc/dn\ .
\label{eq:4.1}
\ee
Let the defect concentration fluctuations be randomly distributed with a second moment 
$\langle\delta n({\bm x})\,\delta n({\bm y})\rangle \approx n\,\delta({\bm x} - {\bm y})$.
The definition of the dimensionless disorder $\delta$ in Eqs.~(\ref{eq:3.2}, \ref{eq:3.3}) then leads to the
estimate $\delta \approx (n/a^3\Tc^2)(d\Tc/dn)^2$. Now model the $n$-dependence of $\Tc$ by
$\Tc \approx \Tc^0 [1 - (n/n_c)^2]$, which is roughly the shape of the phase diagram in, for instance,
many quantum ferromagnets that display a quantum phase transition triggered by chemical composition.\cite{Brando_et_al_2016}.
This leads to an estimate
\be
\delta \approx \frac{1}{a^3n_c}\,(n/n_c)^3\ .
\label{eq:4.2}
\ee
For $n$ a sizable fraction of $n_c$ this is of $O(1)$. The observed strong dependence of $\Tc$ on
a dopant concentration in many systems thus translates into a rather large value of $\delta$.
At least in some materials is therefore would be misleading to conclude, from the fact that they
are good metals, that disorder is irrelevant. We also note that weak-localization effects are not commonly
observed in quantum ferromagnets with a ferromagnet-to-paramagnet quantum phase
transition driven by composition.\cite{Brando_et_al_2016} This is another example of
disorder that couples to the order parameter, but is not easily observed in the transport properties.

We also mention that there was no a priori guarantee that the energetic considerations in Sec.~\ref{sec:II}
would lead to an observable size of the inhomogeneity region for reasonable values of the disorder,
as we saw in Sec.~\ref{sec:III} is indeed the case.
The fact that we found the wedge-shaped region where minority-phase droplets are stable to be
of an observable size is nontrivial and consistent with our interpretation of the existing experiments.

\subsection{The ordered-volume fraction in the ordered and disordered phases}
\label{subsec:IV.B}

In Sec~\ref{subsec:III.B} we have discussed the ordered-volume fraction $F_{\text{OV}}$ as the first-order transition
is approached from the ordered phase. We found that $F_{\text{OV}}$ drops below unity due to
the existence of droplet in a certain region bounded by a dimensionless distance $t_c$ from the coexistence curve.
$t_c$ depends on the strength of the disorder and the strength of the first-order transition, see
Eq.~(\ref{eq:3.15b}). $F_{\text{OV}}$ then decreases monotonically with decreasing $t$. In the
asymptotic region $t < t^*$, with $t^*$ given by Eq.~(\ref{eq:3.14}), some of our simple assumptions
become questionable. For instance, the assumption of noninteracting droplet will certainly break
down with increasing droplet density. It is likely that at some point a percolation transition will lead
to droplets merging, and the precise behavior on or very close to the coexistence curve is a very
hard problem. For instance, it is not obvious whether $F_{\text{OV}}$ is continuous or discontinuous
across the coexistence curve, and more detailed considerations are necessary to determine this.

Another issue is the behavior on the disordered side of the phase transition; we have considered the 
existence of minority-phase droplets in an ordered majority phase only. At first sight one might think
that the behavior should be roughly symmetric with respect to the phase boundary; however, this
is likely not the case. For instance, our estimate of the droplet-wall width $\Ldw$ in Sec.~\ref{subsubsec:III.A.4}
assumes that most of the surface free-energy cost of forming the droplet is paid within the bulk
ordered phase. This is plausible at least at a saddle-point level, as the saddle-point differential
equation for a droplet field configuration is very similar to, say, a square-well problem in quantum
mechanics. If the same is true in the disordered phase, then this introduces an intrinsic asymmetry
into the problem: In the disordered bulk there are no Goldstone modes, the problem is always
Ising-like, and $\Ldw$ is likely substantially smaller than in the ordered phase. This will make
droplet energetically less favorably in the disordered phase than in the ordered one. If there are solutions
for which this is not the case, then one needs to add the requirement $L_c > \Ldw$ to the
conditions for droplet existence, as we have noted above.\cite{droplet_asymmetry_footnote}
In either case, the conditions for droplet existence are more stringent in the disordered phase.
While speculative,
these considerations are consistent with the experimental observations, which generally see
much weaker indications of phase separation in the disordered phase than in the ordered one.
Another contributing factor may be experimental limitations: If droplets are intrinsically smaller
in the disordered phase, then the spatial resolution limit of any experimental technique will 
lead to an underestimation of the ordered-volume fraction.

\subsection{First-order transitions in the presence of disorder}
\label{subsec:IV.C}

The fate of a first-order transition in the presence of disorder is a problem with a long history,
going back to Imry and Ma, and Imry and Wortis.\cite{Imry_Ma_1975, Imry_Wortis_1979}
The relevant arguments have been made rigorous in Refs.~\onlinecite{Aizenman_Wehr_1989,
Aizenman_Wehr_1990} for classical systems, and in Refs.~\onlinecite{Greenblatt_Aizenman_Lebowitz_2009,
Aizenman_Greenblatt_Lebowitz_2012} for the quantum case.\cite{random-field_footnote}
The conclusion of these studies is that no sharp first-order phase transition is possible in
dimensions $d\leq 2$ for systems with a discrete symmetry, and in $d \leq 4$ for systems with
a continuous symmetry. Consistent with this, Fern{\'a}ndez et al found a sharp first-order transition
in a $3$-$d$ disordered Potts model.\cite{Fernandez_et_al_2008}  In the present context these
rigorous results provide, strictly speaking, no constraints: We have considered only systems with
a discrete symmetry (as is always the case for any system on a lattice), so in $d=3$ a sharp
first-order phase transition is possible. One might wonder, however, if in the case of a weakly
broken continuous symmetry any sharp transition must be weakly first order. To see whether
our droplet scenario is consistent with this hypothesis would require a determination of the
ordered-volume fraction on the coexistence curve, where our arguments are not reliable.
We can imagine four scenarios: 
(1) The non-overlapping droplet picture remains valid up to the transition, and $F_{\text{OV}}$ 
(and hence the magnetization) has a discontinuity on the coexistence 
curve. This would mean that there still is a sharp first-order transition. 
(2) The droplets merge at some critical droplet density, and the resulting percolation transition 
is first order,\cite{Cho_Kahng_2015} leading again to a discontinuous $F_{\text{OV}}$.
(3) $F_{\text{OV}}$ changes continuously from unity in the ordered phase to exponentially small 
values in the disordered phase. This would mean no sharp transition, i.e., the original sharp 
first-order transition has been smeared due to the existence of droplets. 
(4) A second-order percolation transition occurs in the droplet system. This would mean there 
still is a sharp transition, but it is continuous rather than first order. 

More detailed work is needed in order to determine which of these possibilities is realized. We 
note that our scenario is entirely consistent with possibilities (1) - (3). If (4) is realized, then some
of our arguments will have to be reconsidered, at least close to the transition, in order to ensure consistency.

We finally comment on the relation between scaling descriptions of first-order phase transitions
\cite{Fisher_Berker_1982, Privman_Fisher_1983} and the rigorous bound for the correlation-length
exponent, Eq.~(\ref{eq:2.1}). The scaling description leads a relevant operator with scale
dimension $\lambda = 1/\nu = d$. Naively, this violates Eq.~(\ref{eq:2.1}) and thus seems to
preclude a sharp first-order transition in any dimension for any nonzero amount of disorder.
This apparent contradiction is resolved by the realization that there are two distinct correlation-length
exponents: One is related to the rounding of the transition in system of finite size $L$ and is equal
to $\nu_{\text{rounding}} = 1/d$ at any regular first-order transition.\cite{Fisher_Berker_1982, RFOT_footnote} The
other is related to the effects of the disorder, which shifts the transition point relative to the clean
one. For uncorrelated disorder, this exponent is equal to $\nu_{\text{shift}} = 2/d$. The theorem proven by
Chayes et al.\cite{Chayes_et_al_1986}, Eq.~(\ref{eq:2.1}) applies to the latter, so there is no
contraction. The presence of two positive exponents is not at odds with the usual notion
of only one relevant operator at a fixed point describing a phase transition, since the tuning
parameter for a first-order transition is $r(L) L^d$, with $r(L) = r_{\infty} + r_1/L^{d/2}$ the 
scale-dependent mass parameter. The tuning parameter is thus $r_{\infty}L^d + r_1 L^{d/2}
= r_{\infty}L^d[1 + (r_1/r_{\infty})/L^{d/2}]$, and the operator with scale dimension $d/2$ describes
corrections to scaling in a well-defined sense.

\acknowledgments
We thank John Toner and Yasutomo Uemura for discussions. This work was supported by the NSF under 
grant Nos. DMR-1401410 and DMR-1401449. 

\vskip 1cm
\appendix

\section{The length scale $\Ldw$ in an Ising model}
\label{app:A}

Consider an action
\be
{\cal A} = {\cal A}^{(2)} + \int_V d{\bm x}\,\left[v\,\phi^3({\bm x}) + u\,\phi^4({\bm x})\right]
\label{eq:A.1}
\ee
with ${\cal A}^{(2)}$ from Eq.~(\ref{eq:3.1}), which allows for a first-order transition by virtue of the cubic
term with coupling constant $v$. In natural units, $u = (J/a^3){\hat u}$, $v = (J/a^3){\hat v}$, with ${\hat u}$
and ${\hat v}$ dimensionless. 
In a mean-field approximation, this model has a first-order transition at $r = r_1 = 2v^2/9u$
where the order parameter changes discontinuously from $\phi = \phi_1 = 2v/3u$ to zero. Now consider a
situation where the mass parameter $r(x,y,z)$ changes discontinuously from $r<r_1$ for $x>0$ to $r>r_1$ for
$x<0$, which models a plane droplet wall, and look for a variational solution of the saddle-point equation
\be
c\,\phi''(x) = r\,\phi(x) - v\,\phi^2(x) + u\,\phi^3(x)
\label{eq:A.2}
\ee
that obeys the boundary conditions $\phi(x=0)=0$ and $\phi(x\to\infty) = {\rm const}$. Deep inside the
ordered phase, for $r$ negative and large, the $v$-term is negligible and the solution will vary on the
microscopic length scale $a$. For $r=0$ the characteristic length scale, which determines the droplet
wall thickness, is $a\,{\hat u}^{1/2}/{\hat v}$, and for $r \to r_1$ it is $3a\,{\hat u}^{1/2}/\sqrt{2}\,{\hat v}$.
The former result is obtained by suitable scaling of the ODE; the latter, by linearizing it about $\phi_1$.
We conclude that upon approaching the coexistence curve, $L_{\text{dw}}$ increases from a $L_{\text{dw}} \approx a$
deep inside the ordered phase to a value on the order of $L_{\text{dw}} \approx a\,{\hat u}^{1/2}/{\hat v}$
in the vicinity of the coexistence curve. If ${\hat u} = O(1)$, we can write this as
\be
\Ldw \approx a/\varphi\ ,
\label{eq:A.3}
\ee
which recovers Eq.~(\ref{eq:3.13}) specialized to the Ising case ($\L_g \approx a$).
For typical parameter values one expects $\Ldw$ to saturate
at a few times the microscopic length. 

\section{Conditions for the existence of droplets within a $\phi^4$-theory}
\label{app:B}

Here we discuss the disorder threshold requirement expressed by Eq.~(\ref{eq:3.15a}) in the framework
of the Ising model action given by Eq.~(\ref{eq:A.1}). In the region $t \alt {\hat r_1} \equiv 2{\hat v}^2/9{\hat u}$,
which corresponds to $r \agt 0$, we have $\phi \approx \varphi_1 = \sqrt{2{\hat r}_1/{\hat u}}$. With
$\Ldw \approx a\,{\hat u}^{1/2}/{\hat v}$ from Appendix \ref{app:A}, Eq.~(\ref{eq:3.15a}) yields
\be
\delta_c \approx {\hat u}^{3/4}\,\varphi_1^{1/2}\ .
\label{eq:B.1}
\ee
Deep inside the ordered phase, ${\hat r} \approx -1$, we have $\phi = \sqrt{-{\hat r}/{\hat u}} \approx 1/{\hat u}^{1/2}$,
and at a moderately strongly first-order transition $\varphi_1$ is, say, one-tenth of that value. The remaining
question is the value of ${\hat u}$. Within Hertz's model for itinerant quantum ferromagnets,\cite{Hertz_1976}
${\hat u} = 1/12$, which leads to $\delta_c \approx 0.1$. A dimensionless disorder strength on this order seems
realistic in the light of the discussion in Sec.~\ref{subsec:IV.A}. For systems with an approximate continuous
symmetry, where $\Ldw$ is larger, the disorder threshold will be correspondingly lower.


\begin{thebibliography}{51}
\bibitem[{dyn()}]{dynamics_footnote}
\bibinfo{note}{Whether or not one introduces the notion of a frequency here is
  a matter of taste. Alternatively, one can define $L_g$ as the length scale
  that cuts off the long-range nature of the correlations that characterized
  the Goldstone modes, without any reference to dynamics.}
\bibitem[{act()}]{action_footnote}
\bibinfo{note}{Our arguments apply equally to thermal and quantum phase
  transitions, even though Eq.~(\ref{eq:3.1}) represents a classical action.
  Generalization to a quantum mechanical action does not affect the following
  discussion.}
\bibitem[{dis()}]{disorder_strength_footnote}
\bibinfo{note}{We note that the dimensionless factors that are left
  undetermined in these purely dimensional arguments are not necessarily close
  to unity. For instance, one can argue that $\rho$ should be proportional to
  $J^2$ times the microscopic wave number $2\pi/a$ cubed, which leads to a
  factor of $(2\pi)^3$ in Eq.~(\ref{eq:3.3}).}
\bibitem[{dro({\natexlab{b}})}]{droplet_asymmetry_footnote}
\bibinfo{note}{We need to ask whether $L_c > L_{\text{dw}}$ is also necessary
  for the existence of well-defined droplets. In the ordered phase this
  condition is {\em not} necessary, since a typical order-parameter
  configuration will decrease gradually as the droplet is approached from the
  outside, and then decrease exponentially inside the droplet. However, for
  droplets to exist in the {\em disordered} phase it may be necessary if the
  order parameter is exponentially small outside the droplet and needs to build
  gradually inside. This is one possible reason for the observed asymmetry
  between the ordered and disordered phases, see the discussion.}
\end{thebibliography}

\end{document}